\documentclass[journal=ascphotonics,manuscript=article]{achemso}

\usepackage{graphicx} 
\usepackage{pdfpages}
\usepackage{amsmath}
\usepackage{amssymb}
\usepackage{ccaption}
\usepackage{nicematrix}
\usepackage{makecell}
\usepackage[version=4]{mhchem}

\author{Eleonora P. Kraus}
\email{eleonora.kraus@physik.uni-marburg.de}
\affiliation{Department of Physics, Philipps-Universit\"at Marburg, D-35032 Marburg, Germany}
\alsoaffiliation{mar.quest | Marburg Center for Quantum Materials and Sustainable Technologies, Hans-Meerwein-Straße 6, D-35032 Marburg, Germany}

\author{Jamie M. Fitzgerald}
\affiliation{Department of Physics, Philipps-Universit\"at Marburg, D-35032 Marburg, Germany}
\alsoaffiliation{mar.quest | Marburg Center for Quantum Materials and Sustainable Technologies, Hans-Meerwein-Straße 6, D-35032 Marburg, Germany}

\author{Carlos Maciel-Escudero}
\affiliation{Department of Physics, Philipps-Universit\"at Marburg, D-35032 Marburg, Germany}
\alsoaffiliation{mar.quest | Marburg Center for Quantum Materials and Sustainable Technologies, Hans-Meerwein-Straße 6, D-35032 Marburg, Germany}

\author{Ermin Malic}
\affiliation{Department of Physics, Philipps-Universit\"at Marburg, D-35032 Marburg, Germany}
\alsoaffiliation{mar.quest | Marburg Center for Quantum Materials and Sustainable Technologies, Hans-Meerwein-Straße 6, D-35032 Marburg, Germany}

\title{Engineering strong coupling in ultra-compact photonic crystal/2D material platforms}

\begin{document}

\begin{abstract}
Sub-wavelength thick photonic crystal (PhC) slabs coupled to 2D excitonic materials, such as transition metal dichalcogenides (TMDs), are a promising platform for highly tunable, room-temperature, on-chip optoelectronic devices. Unlike conventional Fabry-P\'erot microcavities, these compact open cavities exhibit non-trivial electric field profiles, leading to spatially distinct regions of weak and strong coupling with excitons within the PhC unit cell. Using coupled mode theory and rigorous solutions to Maxwell's equations, we investigate how the PhC geometry can be used to control these coexisting exciton/polariton contributions and tailor the resulting optical spectra. For large filling factors, i.e., small air gaps, we show that PhC polaritons can be modeled as dark waveguide modes brightened via the periodicity of the PhC slab. Furthermore, by spatially patterning the TMD monolayer based on the local field intensity, we reveal the simultaneous presence of excitons in both the weak and strong coupling regimes. Overall, this work provides fundamental insights into the strong light-matter coupling regime in structured photonic environments, offering a pathway to design and optimize metal-free, ultra-compact polaritonic devices.
\end{abstract}

\section{Introduction}

Integrating two-dimensional (2D) semiconductors with sub-wavelength thick photonic crystal (PhC) slabs provides a versatile and compact platform for developing on-chip photonic devices \cite{ballarini2013all,pospischil2014solar}. The large exciton oscillator strength in these ultra-thin semiconductors \cite{wang2018colloquium,perea2022exciton} combined with the high-Q factors and small mode volumes of PhC slab optical resonances \cite{hsu2013observation,yoon2015critical} leads to strong light-matter coupling \cite{zhang2018photonic}. This enables the formation of hybrid light-matter quasiparticles known as exciton-polaritons, where the rate of energy exchange between the constituent photon and exciton exceeds the individual decay rates \cite{schneider2018two, ferreira2024}. Conventional polariton devices based on Fabry-P\'erot (FP) cavities made of thick planar distributed Bragg reflectors (DBRs) \cite{liu2015strong,dufferwiel2015exciton} are bulky and offer limited mode control. Furthermore, cavities based on metallic mirrors \cite{wang2016coherent}, plasmonic nanoantennas \cite{kleemann2017strong,wen2017room} and nanoantenna arrays \cite{lee2015fano, timmer2023plasmon} suffer from intrinsically large ohmic and radiative losses, restricting polariton lifetimes to timescales of $\sim 100\,$fs \cite{vasa2013real}. In contrast, all-dielectric PhC slabs can support high-Q factor resonances that, in the case of bound states in the continuum, are limited only by fabrication imperfections and residual material loss \cite{hsu2013observation,hsu2016bound}. The strong spatial confinement of these PhC modes leads to a significant enhancement of the local electromagnetic field \cite{yoon2015critical}, which can interact with a 2D semiconductor embedded in or proximal to the slab \cite{zhang2018photonic}. The resonant wavelengths are primarily determined by the geometry of the structure rather than intrinsic material resonances, meaning they can be tuned to operate over a wide range of energies. In the context of polaritonics, this enables the tailoring of the effective mass and group velocity \cite{xie20252d}, Rabi splitting \cite{gogna2019photonic}, radiative loss \cite{chen2020metasurface}, emission direction and polarization \cite{wang2024polarization,lee2025bound}, and the topological phase \cite{liu2020generation,he2023polaritonic}. Unlike FP cavities with a fixed isotropic dispersion, PhC slabs allow for a more complex energy landscape including highly anisotropic \cite{zhang2018photonic} and flat \cite{do2025room} bands, off-$\Gamma$ minima \cite{chen2020metasurface}, and inverted bands \cite{wang2024polarization,xie20252d} with a negative effective mass \cite{wurdack2023negative}. 

 The integration of 2D semiconductors with PhC slabs offers a route to compact, ultrafast logic devices based on polariton nonlinearities, potentially overcoming the speed limitations of traditional electronics \cite{genco2025femtosecond}. Building on earlier demonstrations using conventional inorganic semiconductor heterostructures \cite{bajoni2009exciton,ardizzone2022polariton} at cryogenic temperatures, recent studies have expanded PhC polaritonics to room temperature using materials such as transition metal dichalcogenides (TMDs) \cite{zhang2018photonic,gogna2019photonic, kravtsov2020nonlinear,chen2020metasurface,he2023polaritonic,maggiolini2023strongly,weber2023intrinsic,qing2020strong}, organics \cite{yan2025topologically} and perovskites \cite{kim2021topological, dang2020tailoring, dang2022realization, wu2024exciton}. Unlike FP cavities that are typically designed for vertical confinement of light, PhC platforms support propagating guided-mode resonances (GMRs) that are better suited for efficient exciton-polariton transport \cite{dang2024long, xie20252d, fitzgerald2025polariton,ferreira2022microscopic}. Understanding how PhC design influences the coupling strength with 2D semiconductors is therefore crucial for optimizing light-matter interactions in these hybrid photonic systems. In contrast to FP cavities, where transmission of light can only occur via interaction with a polariton \cite{fitzgerald2022twist}, PhC slabs support both direct and indirect pathways \cite{fan2003temporal}. The interference between these leads to the conventional asymmetric Fano lineshape of the leaky GMR in reflection and transmission \cite{miroshnichenko2010fano}. Furthermore, direct and indirect pathways result in optical spectra featuring three distinct peaks, which have been observed in experiments involving plasmonic platforms combined with organic materials \cite{salomon2013strong,vasa2013real,todisco2015exciton,timmer2023plasmon}, TMDs\cite{vadia2023magneto}, self-hybridized van der Waals metasurfaces \cite{weber2023intrinsic}, and nanodisks \cite{verre2019transition}. This triplet structure has been attributed to strongly and weakly coupled excitons \cite{nguyen2022silicon,greten2024strong}, but a comprehensive theoretical approach explaining the origin of this third peak is still lacking for dielectric PhC systems. 

In this theoretical work, we investigate the strong coupling between excitons in a representative \ce{WS2} monolayer \cite{wang2018colloquium, mueller2018exciton} and GMRs of a PhC grating slab \cite{fan2002analysis}. By solving the full Maxwell's equations with realistic material parameters, we explore how the characteristics of PhCs impact strong light-matter coupling. Using coupled-mode theory (CMT) \cite{fan2003temporal}, we include the spatially separated weakly and strongly coupled excitons within the PhC unit cell to accurately model the spectral position and intensity of the three peaks appearing in optical spectra. Additionally, by patterning the monolayer based on the high- and low-field regions of the unit cell, we explicitly demonstrate the presence of both weakly and strongly coupled excitons. This approach provides control over the Rabi splitting and the multi-peak structure in the polariton absorption spectra. Furthermore, we show that the coupling strength can be estimated from the electric field and the dielectric profile of the PhC slab using the mode volume and electric field intensity at the position of the TMD monolayer. Overall, our work provides a thorough understanding of the strong coupling regime in ultra-thin PhC/2D material platforms by linking their optical response to the underlying electromagnetic mode profile of PhC gratings. These insights can offer guidance for the development of polariton-based optoelectronic devices, combining weakly and strongly coupled exciton physics on the scale of tens of nanometers.

\section{Photonic crystal characterization}

\begin{figure}[ht!]
    \centering
    \includegraphics[width=0.85\textwidth]{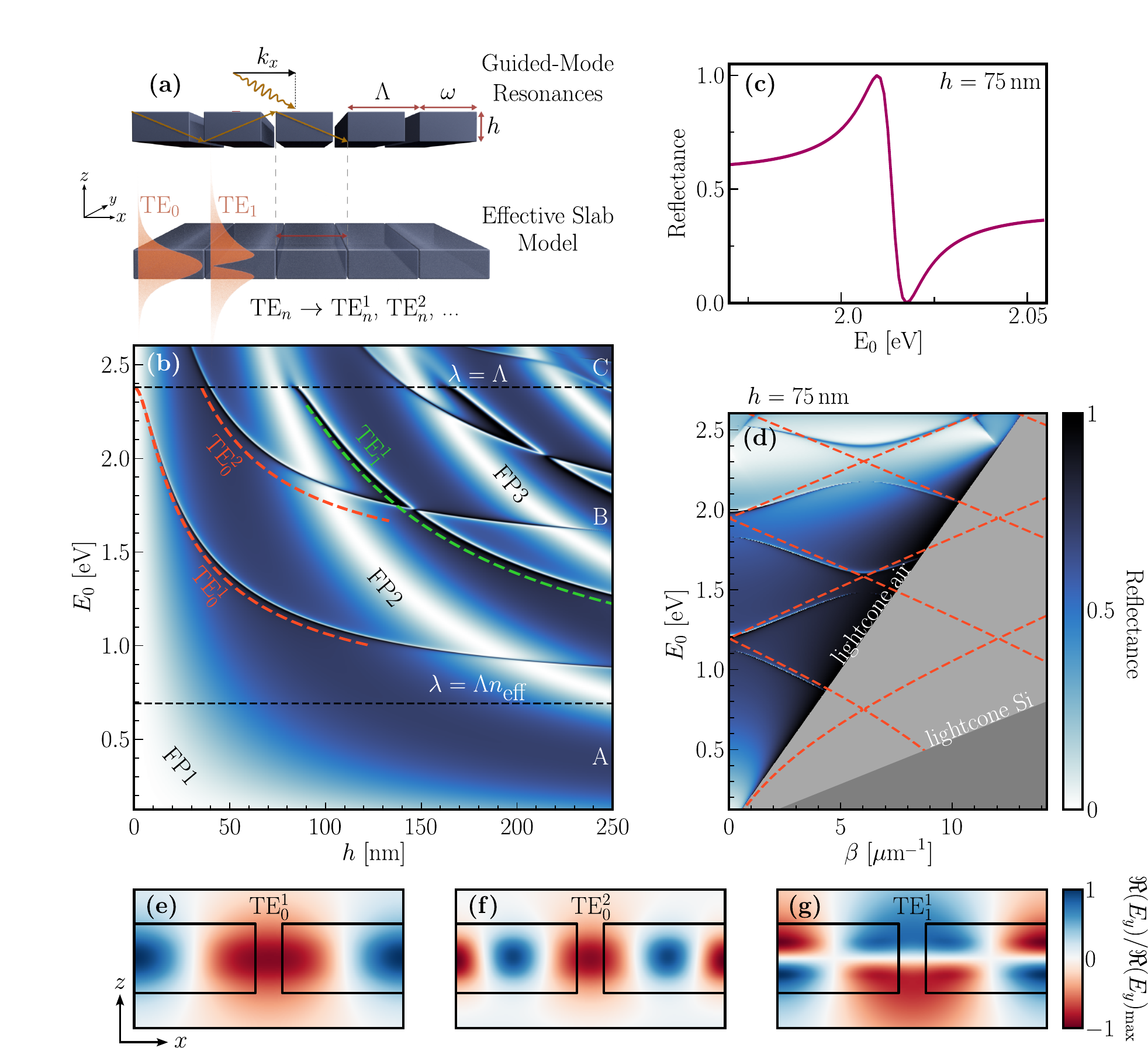}
    \caption{\textbf{(a)} Schematic of a 1D PhC slab with thickness $h$, period $\Lambda$, ridge width $w$, and incident light (orange arrow) with an in-plane momentum $k_x$. Below is a representation of the effective slab model used to characterize the guided-mode resonances (GMRs). \textbf{(b)} Reflectance spectra of the bare grating at normal incidence versus energy $E_0$ and thickness $h$ for a fixed $\Lambda = 521\,$nm, $w = 468.9\,$nm, and filling factor $f= w/\Lambda =0.9$. Horizontal dashed lines mark three distinct regimes: (A) Fabry-P\'erot ($\lambda > n_{\text{eff}}\Lambda$), (B) guided-mode ($n_{\text{eff}} \Lambda >\lambda > \Lambda$), and (C) diffractive ($\lambda<\Lambda$). The orange and green dashed lines indicate the TE$_0^{1}$/TE$_0^2$ and TE$_1^1$ GMR dispersions, respectively, calculated with Eqs.~\ref{eq:s_zeros} and \ref{eq:kz}. The broad background Fabry-P\'erot modes of the spectra are labeled as FP1 = TE$_0^0$, FP2 = TE$_1^0$, etc. \textbf{(c)} Reflectance spectrum of the TE$_0^2$ mode at $h = 75$\,nm, revealing the characteristic asymmetric Fano lineshape. \textbf{(d)} Reflectance spectra of the bare grating versus energy $E_0$ and in-plane momentum $\beta$. Light-gray shaded area indicates the region that supports slab waveguide modes $E_0/(c_0\hbar) < \beta < n_{\text{Si}}E_0/(c_0 \hbar)$. The dispersion of the first diffraction order TE$_0$ mode, folded back into the external lightcone, is indicated by the orange dashed lines. \textbf{(e)-(g)} Normalized real part of the $y$-component of the electric field distribution in a single unit cell for the TE modes labeled in (b). All colormaps are calculated using RCWA.}
    \label{fig:big_sweep}
\end{figure}

In this work, we focus on sub-wavelength thick silicon-based ($n=3.52$) one-dimensional (1D) PhC slabs, specifically near-wavelength high-contrast lamellar gratings, as shown in the top panel of Fig.~\ref{fig:big_sweep}a. The ridge width $w$, lattice period $\Lambda$, and thickness $h$ provide three independent parameters to tune the optical response of the PhC slab. The material loss of silicon is neglected. All simulations presented below were performed using rigorous coupled wave analysis (RCWA), a semi-analytical method ideal for efficiently modeling layered structures with in-plane periodicity \cite{whittaker1999scattering}. Throughout, we excite the 1D grating such that the modes supported by the PhC propagate perpendicular to the ridges ($x$-axis), rather than along the invariant $y$-axis.

Figure~\ref{fig:big_sweep}b shows the reflectance spectra of the PhC for typical parameters of $\Lambda = 521\,\mbox{nm} , h = 72.5\,\mbox{nm}$ and $w = 0.9\Lambda$. The optical response can be split into three distinct spectral regions \cite{karagodsky2012physics}: (A) The deep-subwavelength Fabry-P\'erot regime ($\lambda>\Lambda n_{\text{eff}}$), where the incident wavelength is much larger than the PhC period. In this limit, the grating acts as an effective uniform dielectric slab with the refractive index $n_{\text{eff}}$, supporting only FP slab modes (white broad bands of low reflectance in Fig.~\ref{fig:big_sweep}b). (B) The intermediate near-wavelength regime ($\Lambda <\lambda<\Lambda n_{\text{eff}}$) exhibits well-defined, sharp GMRs \cite{quaranta2018recent}, labeled as TE$_n^m$ according to their spatial field profile as described below. Here, the structure supports internally propagating diffractive orders while remaining non-diffractive in the far-field, i.e., the resonant modes couple only to the zeroth diffraction order in the air region. (C) The diffractive regime, where the incident wavelength is much smaller than the period of the grating ($\lambda<\Lambda$) and higher diffraction orders can be observed in the reflectance spectra \cite{karagodsky2012physics}. The sharp features in region (B) are the consequence of leaky GMRs. At a specific incident angle and wavelength, light can be coupled into the waveguide modes supported by the slab via diffraction orders \cite{quaranta2018recent}. The reverse process is also possible, i.e., the waveguide modes can couple to the far-field via the grating. This radiative leakage interferes with the directly reflected or transmitted light of the broad FP mode, leading in general to sharp asymmetric Fano resonances in reflection and transmission spectra \cite{fan2003temporal}, as shown in Fig~\ref{fig:big_sweep}c. In contrast to FP cavities, the spectral position of the GMRs in the PhC can be tuned to the minima or maxima of the background FP slab mode reflectance, as shown in Fig~\ref{fig:big_sweep}b. This offers a great degree of spectral flexibility for non-resonant exciton-polariton excitation, whereas in FP microcavities, non-resonant pumping must occur outside the cavity stopband.

In the case of thin PhC gratings and large filling factors, $f$, the leaky GMRs in region (B) can be well described by combining a slab waveguide model with the diffraction grating equation \cite{chen2017high, quaranta2018recent}. Within this approach, which is referred to as the effective slab model in this work, the PhC slab can be considered as a thin, effective, homogeneous slab that supports both transverse electric (TE) and magnetic (TM) waveguide modes. We label these as TE$_n$ and TM$_n$, respectively, where $n$ is the number of nodes along the vertical ($z$-) direction. In addition, the impact of the grating periodicity is approximated as a periodic perturbation that imparts an in-plane momentum to couple slab waveguide modes to the far-field. This empty lattice approximation \cite{tikhodeev2002quasiguided} is depicted in the bottom panel of Fig.~\ref{fig:big_sweep}a. In this case, the propagation wavevector of the waveguide mode is set equal to the in-plane wavevector of the incident light plus the momentum imparted by the grating, that is $\beta = k_x + m 2\pi/\Lambda$ where $m\in \mathbb{Z}$ is the diffraction order. The condition for a slab waveguide mode is given by the round-trip phase accumulation \cite{chen2017high}
\begin{equation}
    \label{eq:s_zeros}
    1 + r_{12}r_{23} e^{2i k_z^{(2)} h} = 0,
\end{equation}
where $r_{ij}^{\text{TE}}=(k_z^{(i)} - k_z^{(j)})/(k_z^{(i)} + k_z^{(j)})$ and $r_{ij}^{\text{TM}}=(n_j^2 k_z^{(i)} - n_i^2 k_z^{(j)})/(k_z^{(i)} + k_z^{(j)})$ are the Fresnel coefficients for the interface between the dielectric media $i$ and $j$. The out-of-plane momentum component is then written as 
\begin{equation}
    k_z^{(2)} =\sqrt{  \left(\frac{\omega}{c_0}n_{\text{eff}}^{(i)}\right)^2 - \left( m \frac{2\pi}{\Lambda} \right)^2},
    \label{eq:kz}
\end{equation}
where $c_0$ is the speed of light in free space and $\omega$ is the angular frequency. In Eq.~\ref{eq:kz}, the effective refractive index of the PhC slab is calculated using the volume average \cite{hemmati2020applicability,lalanne1997depth}, $n_{\text{eff}}=\sqrt{\varepsilon_{\text{Si}}f+\varepsilon_{\text{Air}}(1-f)}$ (see SI for more details). The model can be understood as follows: In a homogeneous dielectric layer, the dispersion of the slab waveguide modes lies inside the Si lightcone (light gray area in Fig.~\ref{fig:big_sweep}d). These are bound modes confined in the slab that decay exponentially outside. Introducing a periodic perturbation zone-folds the slab waveguide dispersion into the air lightcone (see orange dashed lines in Fig.~\ref{fig:big_sweep}d). This couples these momentum-dark modes with the external photon continuum, leading to leaky GMRs. The resulting dispersions from Eqs.~\ref{eq:s_zeros} and \ref{eq:kz} are in good agreement with the RCWA simulations (see dashed lines in Fig.~\ref{fig:big_sweep}b and d). Note that the empty lattice approximation does not capture bandgap openings at the center and borders of the first BZ; it treats the grating periodicity strictly as a mechanism for momentum transfer, neglecting the coupling between different PhC modes \cite{tikhodeev2002quasiguided}.


Within the effective slab model, we characterize each mode as TE$_n^m$/TM$_n^m$, where the electric/magnetic field vector is oriented parallel to the grooves of the grating. Here, the waveguide number $n$ again represents the number of nodes along the vertical direction, and the diffraction order $m$ dictates half the number of nodes along the horizontal ($x$-) direction. This can be observed in Figs.~\ref{fig:big_sweep}e-g for TE$_0^1$, TE$_0^2$ and TE$_1^1$, respectively, where the appearance of electromagnetic hot spots is apparent. These facilitate simultaneous weak and strong light-exciton coupling within the unit cell, which we investigate in the following sections by integrating a TMD monolayer on top of the PhC.

\section{Exciton-polaritons in PhC/TMD systems}
For the excitonic medium, we chose a representative \ce{WS2} monolayer. This 2D material supports stable excitons with binding energies in the range of hundreds of meV and large oscillator strengths \cite{wang2018colloquium,perea2022exciton,berghaeuser2018}. To model the optical response of the monolayer, we use a multi-Lorentzian model with the parameters reported in Ref.~\citenum{li2014measurement} for the room-temperature dielectric function. Our results are general and applicable to other 2D materials, such as layered perovskites \cite{kim2021topological, dang2020tailoring, dang2022realization, wu2024exciton} and different TMD monolayers (e.g., \ce{MoTe2} for near-infrared operation \cite{han2025infrared}), with suitably scaled PhCs. As the TMD monolayer thickness ($d_{\ce{WS2}} = 0.7\,$nm) is negligible compared to that of the grating, the mirror-symmetry of the system with respect to the PhC plane is approximately preserved, and the change in dielectric background due to the TMD monolayer does not significantly alter the PhC modes. We focus on the TE$_0^2$ mode (Fig.~\ref{fig:big_sweep}f), which is well separated from neighboring modes and is resonant with the \ce{WS2} A-exciton for the chosen PhC parameters, specifically $h = 60-85\,$nm.

\begin{figure}[t!]
    \centering
    \includegraphics[width=0.85\columnwidth]{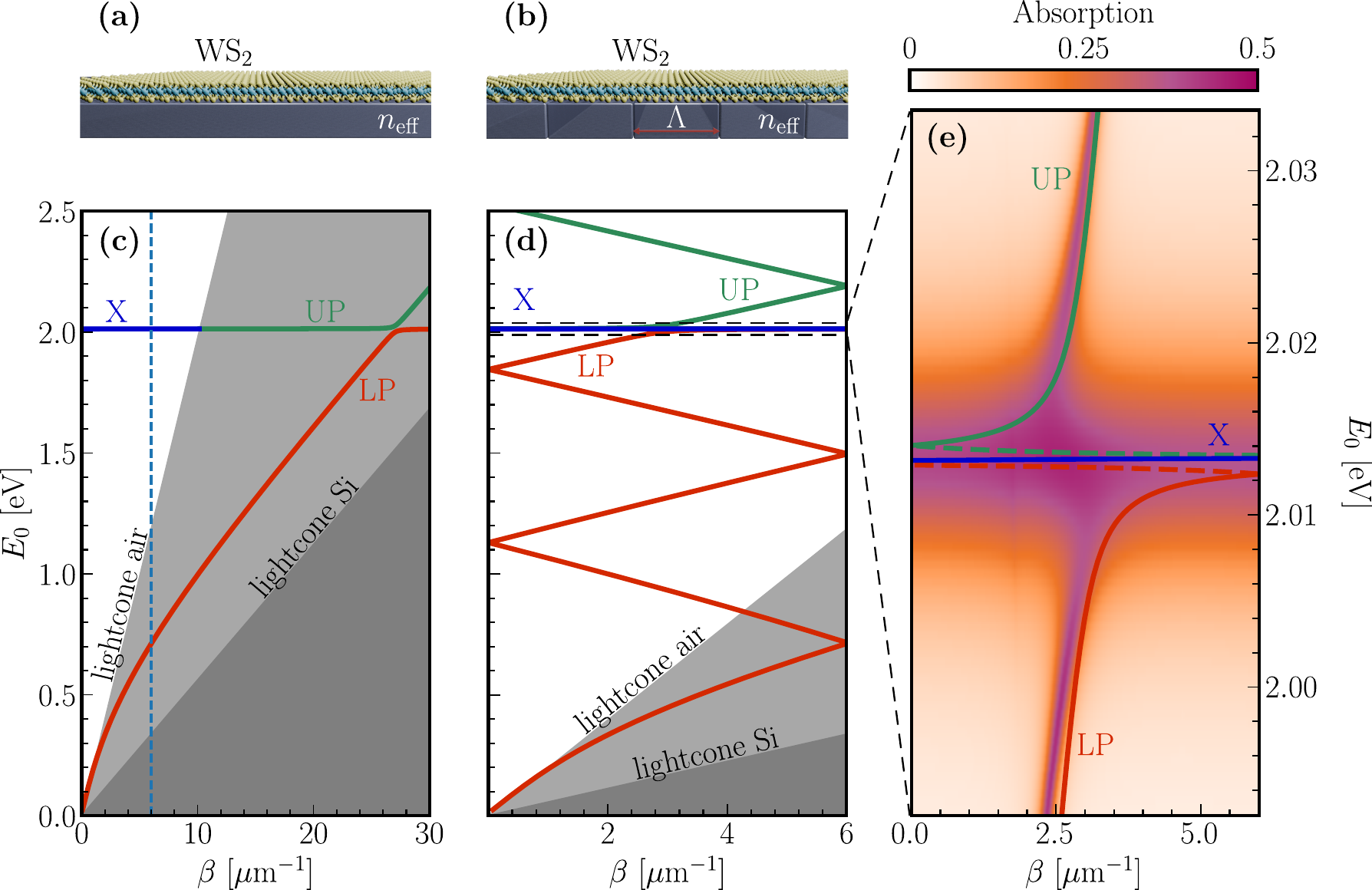}
    \caption{\textbf{(a)} Schematic of the effective homogeneous slab (refractive index $n_{\text{eff}}$) and the \ce{WS2} monolayer. \textbf{(b)} Schematic of the effective homogeneous slab with a periodic perturbation, $\Lambda$. \textbf{(c)} Dispersion of the upper (green line) and lower (red line) slab waveguide polaritons resulting from the hybridization of the TE$_0$ slab waveguide mode and the exciton (blue horizontal line). Without any periodicity, the TE$_0$ mode lies within the silicon lightcone (light-gray area). Only the bare excitons within the air lightcone are bright. The vertical dashed line represents the edge of the 1st BZ, $\beta = \pi / \Lambda$.  \textbf{(d)} Band structure of the guided polaritons resulting from slab waveguide polaritons zone-folded into the 1st BZ within the empty lattice approximation. Both polariton branches are folded into the air lightcone, leading to the brightening of slab waveguide polaritons. Only the positive half of the 1st BZ is shown ($\beta \geq 0$). \textbf{(e)} Close-up of the avoided crossing in (d) comparing the absorption spectra obtained with RCWA simulation (colormap) to the effective slab model (solid lines). Dashed green/red lines indicate the zone-folded UP/LP dispersions from neighboring BZs.}
    \label{fig:polariton_folding}
\end{figure}

The strong interaction between the TE$_0^2$ GMR and the A-exciton resonance results in two polariton branches, which we refer to as guided polaritons. To describe these guided polariton modes within the effective slab model, we consider the strong coupling between the exciton resonance and the slab waveguide mode TE$_0$ (Eq.~\ref{eq:s_zeros}) supported by the effective slab, as illustrated in Fig.~\ref{fig:polariton_folding}a. This approach allows us to calculate the coupling strength as \cite{ciers2017propagating}
\begin{equation}
    \label{eq:g}
    g= \hbar \mathcal{F} \sqrt{\frac{e^2f}{m_\text{e}\varepsilon_0n_{\text{eff}}^2L_{\text{mode}}}},
\end{equation}
where $f$, $L_{\text{mode}}$ and $m_{\text{e}}$ are the exciton oscillator strength per unit area (see SI), the mode length of the slab waveguide mode in the effective slab, and the free electron mass, respectively.  Additionally, the factor $\mathcal{F}$, which accounts for displacement of the TMD monolayer away form the electric field maximum, is defined by the ratio between the electric field at the center of the PhC and the field at the position of the TMD monolayer. Analogous to the earlier interpretation of the GMRs discussed in Fig.~\ref{fig:big_sweep}d, the effect of the grating periodicity (illustrated in Fig.~\ref{fig:polariton_folding}b) considered within the empty lattice approximation is to zone-fold the slab waveguide polariton dispersion into the air lightcone, leading to radiative guided polariton modes. We provide further details and demonstrate the accuracy of this model below. 

Starting with a homogeneous effective slab (Fig.~\ref{fig:polariton_folding}a), the dispersion of the lowest energy slab waveguide mode supported by the structure lies inside the Si lightcone and outside the air lightcone (light-gray area in Fig.~\ref{fig:polariton_folding}c). Despite being momentum-dark, the slab waveguide mode strongly couples to the exciton, which has an approximately flat dispersion for the in-plane momenta on the order of $10\,\mu$m$^{-1}$ (horizontal blue line). The coupling strength $g=13.1\,$meV is calculated using Eq.~\ref{eq:g}. The resulting slab waveguide polaritons also lie inside the Si lightcone and are consequently dark. When imparting an in-plane momentum through a periodic perturbation, capturing the effect of the grating trenches (see schematics in Fig.~\ref{fig:polariton_folding}b), the polariton dispersion is zone-folded into the air lightcone as shown in Fig.~\ref{fig:polariton_folding}d. In Fig.~\ref{fig:polariton_folding}e, we compare these guided polariton dispersions with absorption spectra obtained with RCWA simulations and find excellent agreement between the two without the need of fitting the spectra. 

In the effective slab model, the bright exciton remains unaffected after zone-folding, consistent with the three peaks appearing in the RCWA absorption spectra. In Fig.~\ref{fig:polariton_folding}e, the green/red dashed lines represent the upper and lower polariton branches (UP, LP) zone-folded from neighboring BZs (see SI). The effective slab model does not capture the photonic band gap opening at the edges of the BZ, as already observed in Fig.~\ref{fig:big_sweep}d. In an improved model that incorporates a periodically varying dielectric function, these zone-folded dispersions would be overlaid with the flat exciton dispersion. Conversely, changing the order of our approach — first introducing the periodic perturbation (zone-folding) and then including the strong coupling — correctly predicts the appearance of the energy band gap, but fails to capture the bright exciton resonance in the absorption spectra. Our model has the distinct advantage that, for a suitable choice of $n_{\text{eff}}$, the coupling strength can be explicitly calculated. We use the volume average of the refractive indices of the grating trench and ridge, which for high filling factors, is expected to provide a reasonable estimate of the Rabi splitting. The applicability of this model relies on the accuracy of $n_{\text{eff}}$, which could be further improved with more advanced effective medium theories \cite{hemmati2020applicability,lalanne1997depth} (see SI).


\section{Coexistence of weakly and strongly coupled excitons}
Despite the accuracy of the effective slab model for describing the guided polariton dispersion (Fig.~\ref{fig:polariton_folding}), it cannot provide linear optical spectra. To elucidate the origin of the middle peak at the exciton energy and its impact on the absorption spectra of the structure, we thus employ temporal CMT \cite{fan2003temporal}. This provides the spectral positions, linewidths, and intensities of all resonances. Given the sharply varying electric field profile within a single PhC unit-cell (as shown in Fig.~\ref{fig:big_sweep}e-g), we distinguish two distinct exciton modes: (i) excitons located in regions of high electric field intensity (strongly coupled, SC), which are directly coupled to the photonic mode with a coupling strength given by Eq.~\ref{eq:g}; and (ii) excitons located in the weak-field regions (weakly coupled, WC), which interact dissipatively with the GMR \cite{wurdack2023negative} via interference in the far field. This is an approximation that does not account for the transition region between low- and high-field regions, however the model accurately captures the relevant physics of the system.

Using CMT, we construct the following set of four coupled equations to describe the dynamics of the excitonic and photonic mode amplitude vector $M$, and the input/output field amplitudes $b^{\pm}$:
\begin{equation}
    \begin{split}
        \hbar\partial_t M &= \left(\hbar\Omega + \Gamma\right)M + K^T b^{+} \\
        b^- &= \mathcal{C}b^+ + KM,
    \end{split}   
    \label{eq:CMT}
\end{equation}
where $\mathcal{C}$ represents the scattering matrix of the background process, incorporating the FP slab modes (Fig.~\ref{fig:big_sweep}) and the background permittivity of the TMD. The column vector $M = \left( A_\text{GMR},  X_{sc},  X_{wc}\right)^T$ comprises the TE$_0^2$ GMR amplitude $A_\text{GMR}$ and the SC/WC exciton mode amplitudes $X_{\text{sc/wc}}$. Each mode has a corresponding energy $\hbar\omega_i$ and non-radiative loss $\gamma_{i,\text{nr}}$, while there is a coherent coupling $g$ between the photonic mode and SC excitons, all entering in the matrix $\hbar\Omega$. The matrix $\Gamma$ contains the radiative decays $\gamma_{i,\text{r}}$ and  the dissipative coupling \cite{suh2004temporal} $\tilde{g} = i\sqrt{\gamma_{\text{GMR},\text{r}} \gamma_{X,\text{r}}}$. Taken altogether, we obtain
\begin{equation}
   \hbar\Omega = \begin{pmatrix} \hbar \omega_\text{GMR} && g && 0 \\ g && \hbar \omega_X - i\gamma_{X,\text{nr}} && 0 \\ 0 && 0 && \hbar \omega_X - i\gamma_{X,\text{nr}}\end{pmatrix} ,\qquad \Gamma = i\begin{pmatrix} -\gamma_{\text{GMR},\text{r}} && 0 && \tilde{g}  \\ 0 && 0 && 0 \\  \tilde{g} && 0 && -\gamma_{X,\text{r}}  \end{pmatrix}.
   \label{eq:omega_gamma}
\end{equation}    
 The coupling of the modes to the incoming/outgoing ports $b^\pm$ is governed by
\begin{equation}
     K = \begin{pmatrix} \sqrt{\gamma_{\text{GMR},\text{r}}} && 0 && \sqrt{\gamma_{X,\text{r}}} \\ \sqrt{\gamma_{\text{GMR},\text{r}}} && 0 && \sqrt{\gamma_{X,\text{r}}} \end{pmatrix}. 
\end{equation}
\begin{figure}[t!]
    \centering
    \includegraphics[width=0.6\columnwidth]{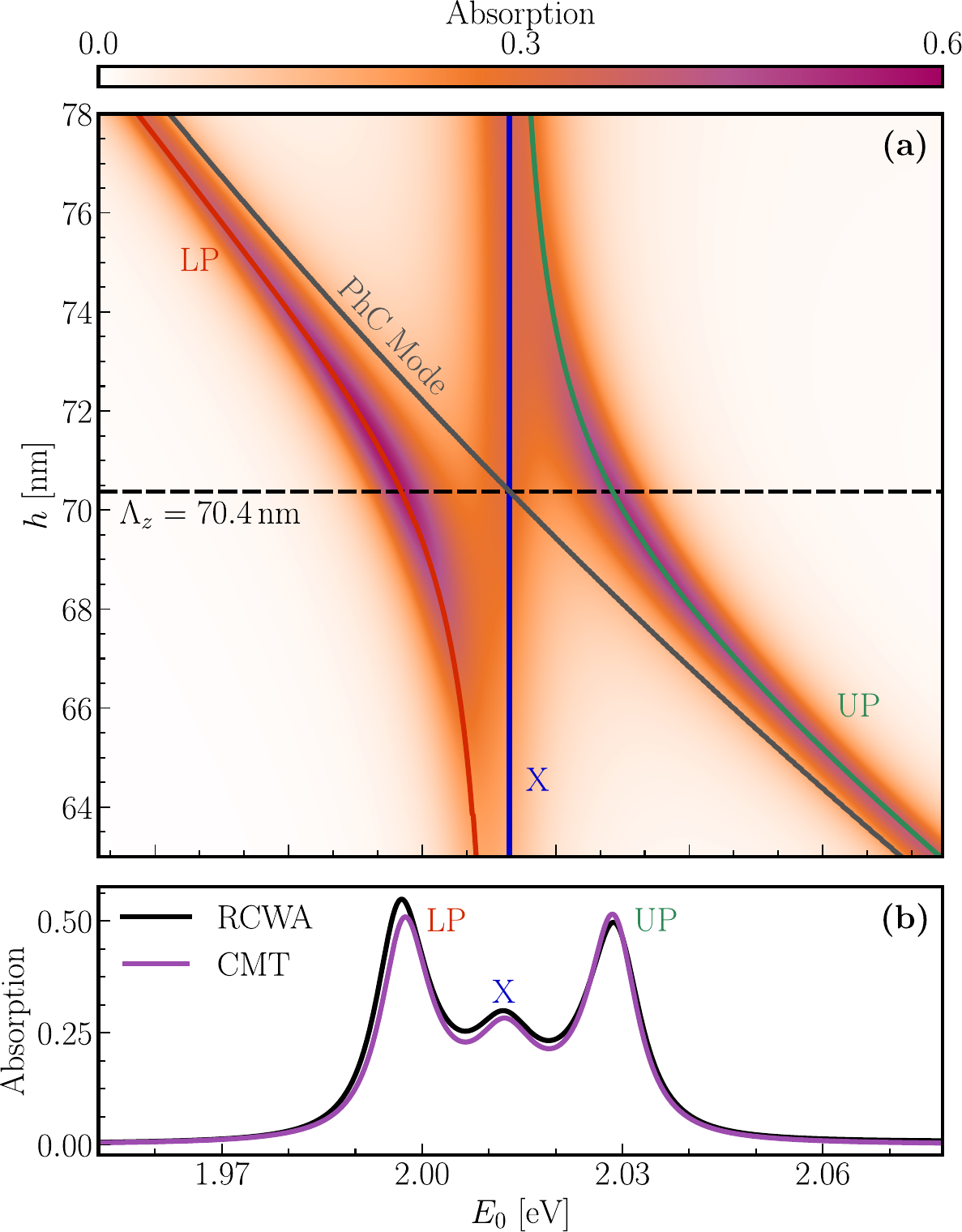}
    \caption{\textbf{(a)} Absorption spectra of the PhC/WS$_2$ structure calculated using RCWA and plotted versus energy $E_0$ and thickness $h$ for a fixed $\Lambda =521$\,nm and $f = 0.9$. A pronounced Rabi splitting of $\hbar \Omega_\text{R}\approx 32$\,meV is observed, alongside the emergence of an additional third peak (blue line), whose spectral position close to the bare exciton energy is approximately independent of $h$. The polariton energies (red/green lines) are calculated using coupled mode theory (CMT), with the coupling strength derived from the effective slab model (Eq.~\ref{eq:g}). \textbf{(b)} Absorption spectrum (solid black line) extracted from the horizontal dashed line in (a) at $h = 70.4\,$nm, highlighting the third peak between the upper and lower polariton peaks. The spectrum is compared to the results obtained with CMT (solid purple line).}
    \label{fig:absorption}
\end{figure}
This matrix describes the coupling of the two radiative modes ($A_\text{GMR}$ and $X_{\text{wc}}$) to the system's ports (corresponding to the radiative continuum in the half-spaces above and below the slab), under the assumptions of mirror symmetry, reciprocity, and time-reversal symmetry \cite{fan2003temporal}. Transforming Eq.~\ref{eq:CMT} into the frequency domain yields the total scattering matrix of the structure, which gives access to linear optical spectra, as well as the polariton energies
\begin{equation}
    \hbar \omega_{P\pm} = \frac{\hbar\omega_X + \hbar\omega_\text{GMR}}{2} + i \frac{\gamma_{X} + \gamma_\text{GMR}}{2} \pm \frac{1}{2}\underbrace{\sqrt{4g^2 + \left[\hbar \omega_X - \hbar\omega_\text{GMR} + i\left(\gamma_{X} - \gamma_\text{GMR}\right)\right]^2}}_{\hbar\Omega_{\text{R}}} 
    \label{eq:polariton_energy}, 
\end{equation}
with the Rabi splitting $\hbar\Omega_{\text{R}}$ and the total dissipation $\gamma_i = \gamma_{i,\text{r}}+\gamma_{i,\text{nr}}$, which corresponds to half of the full width at half maximum of the resonances. We emphasize that the coupling strength $g$ can be determined with Eq.~\ref{eq:g}, eliminating the need for fitting. The key assumptions of this model are: (i) The direct coupling of SC excitons to the ports is neglected as their optical response is much weaker than that of the PhC mode to which they are coupled. This is consistent with the effective slab model, where non-radiative excitons outside the air lightcone strongly couple to the slab waveguide mode, and the resulting guided polaritons are brightened by the grating periodicity (Fig.~\ref{fig:polariton_folding}). (ii) We neglect the non-radiative decay of the GMR, $\gamma_{\text{GMR},\text{nr}}=0$. (iii) The coupling matrix $\hbar\Omega$ in Eq.~\ref{eq:omega_gamma} excludes direct coupling between the WC exciton and the photonic mode, as this is implicitly captured via the Purcell effect in $\gamma_{X,\text{r}}$ (see SI for further discussion).

To investigate the central peak in greater detail, we calculate the absorption spectra of the PhC/\ce{WS2} structure over a range of PhC thicknesses $h$ using RCWA, see Fig.~\ref{fig:absorption}. Consistent with the $\beta$-sweep in Fig.~\ref{fig:polariton_folding}e, we observe three distinct peaks in the spectra, the upper and lower polariton resulting from strong coupling between SC excitons and the GMR, and a third middle peak close to the bare exciton energy. This central peak position exhibits a weak dependence on the PhC thickness (blue line), while the linewidth broadens with increasing $h$. These effects are attributed to the Lamb shift and Purcell effect, respectively, arising from the weak coupling of WC excitons with the GMR. The middle peak is often obscured in reflectance as the Fano-like lineshape diminishes the readability of the spectra. In absorption, however, it is clearly visible as observed in Fig.~\ref{fig:absorption}b. Using the CMT discussed above, we can reproduce the polariton energies (solid lines in Fig.~\ref{fig:absorption}a) and absorption spectrum (purple line in Fig.~\ref{fig:absorption}b) finding an excellent agreement. The three-peak structure in the absorption spectra can be attributed to coexistence of strongly and weakly coupled excitons in the highly non-uniform electric field distribution within the PhC unit cell \cite{vasa2013real, nguyen2022silicon,timmer2023plasmon}. The accuracy of the CMT result further supports the claim of two distinct exciton species. We attribute the asymmetry in the RCWA spectra (Fig.~\ref{fig:absorption}b) to the slight breaking of vertical mirror symmetry due to the presence of the monolayer on top of the PhC. This is further supported by the fact that the maximum LP absorption is larger than $0.5$, which is only possible in asymmetric systems \cite{piper2014total}. This is not accounted for in the CMT as we assume a symmetric decay of the modes into the two ports (Eq.~6), thus limiting maximum absorption to $0.5$.

\begin{figure}[t!]
    \centering
    \includegraphics[width=0.85\textwidth]{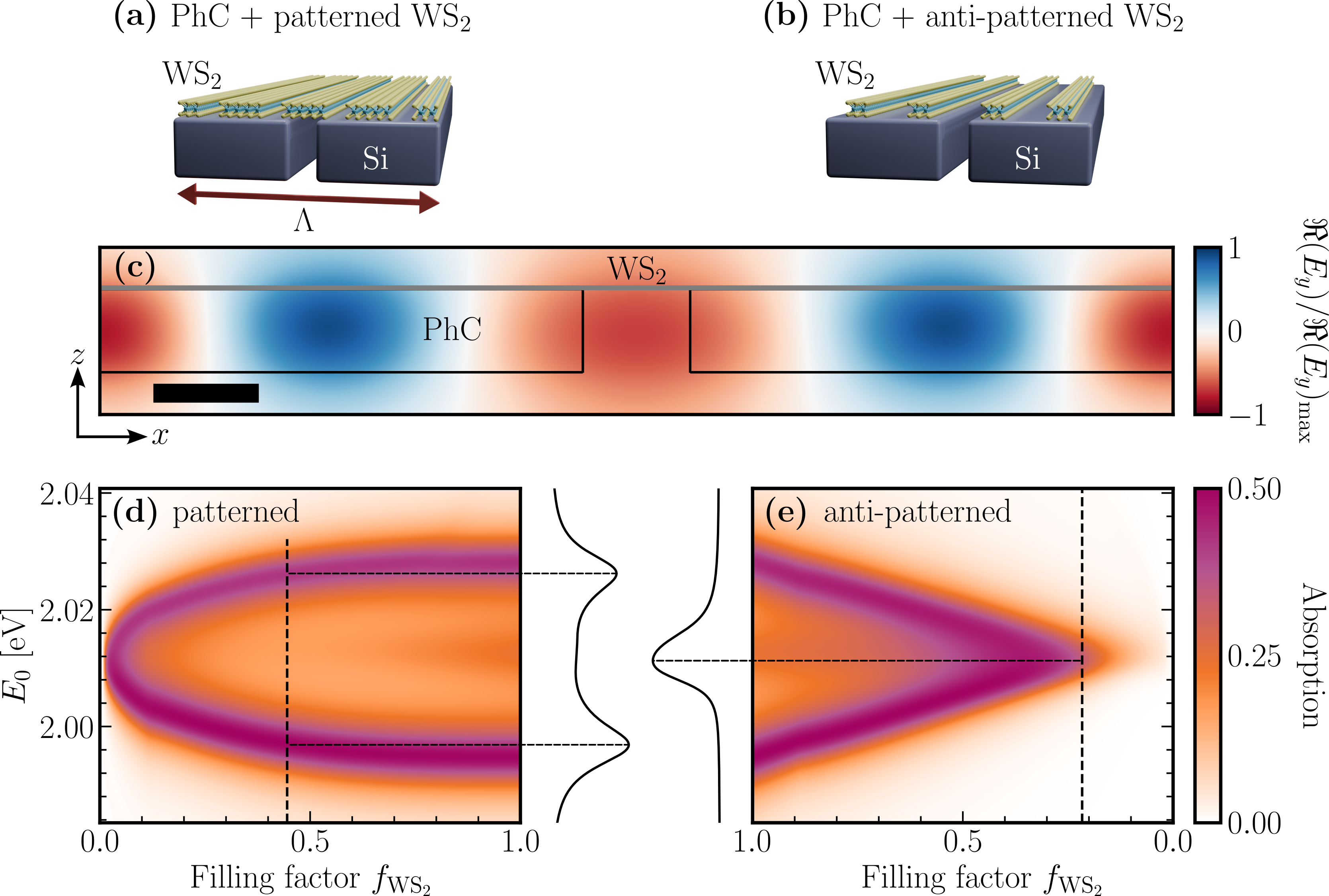}
    \caption{Schematics of the \textbf{(a)} patterned and \textbf{(b)} anti-patterned \ce{WS2} monolayer, where the TMD is removed from regions of low and high electric field intensity, respectively. \textbf{(c)} Normalized in-plane electric field distribution across a single PhC unit cell ($\Lambda = 521$\,nm, $f=0.9$). The electric field of the TE$_0^2$ mode extends into the TMD monolayer, subjecting it to drastically different field intensities in spatially separated regions. The black scale bar is $50$\,nm wide and $20$\,nm high. \textbf{(d)-(e)} Normal incidence absorption spectra of the patterned/anti-patterned structure versus \ce{WS2} filling factor and energy. The two linecuts (solid black lines) in (d) and (e) were extracted from the absorption spectra at $f_{\ce{WS2}}=0.45$ and $0.24$, respectively (vertical dashed lines). All colormaps were calculated with RCWA.}
    \label{fig:patterned}
\end{figure}

To explicitly demonstrate the existence of two spatially distinct exciton species, we simulate the PhC geometry from Fig.~\ref{fig:absorption}a, but now with a patterned \ce{WS2} monolayer. This approach allows us to isolate the effect of high and low electric field regions. We define two configurations: the 'patterned' structure, where the TMD is selectively removed from regions of low electric field intensity (Fig.~\ref{fig:patterned}a); and the 'anti-patterned' structure, where the TMD is removed from regions of high intensity (Fig.~\ref{fig:patterned}b). Analogous to the PhC filling factor $f_{\text{PhC}}$, the \ce{WS2} filling factor, $f_{\ce{WS2}}$, is the fraction of the unit cell covered by the monolayer. In Fig.~\ref{fig:patterned}c, we show the electric field distribution of the TE$_0^2$ mode (same as Fig.~\ref{fig:big_sweep}f), highlighting regions of high and low electric field intensities. Note that in order to minimize the change of the dielectric background of the system, and therefore the direct process, the removed TMD sections were replaced by a weakly dispersive material with the refractive index of the \ce{WS2} background. Recent advancements in the fabrication of monolayer-TMD nanoribbons, i.e., TMD-based 1D PhCs with widths in the range of $500-1000\,$nm \cite{wang2019etching,zhou2022non}, suggest the experimental feasibility of this approach.

Figures~\ref{fig:patterned}d and~\ref{fig:patterned}e show the absorption spectra of these two systems as a function of $f_{\ce{WS2}}$. By removing the TMD from the low-field regions (patterned case), the middle peak in absorption vanishes, recovering the conventional two-peak structure characteristic of strongly coupled systems, see Fig.~\ref{fig:patterned}d at small $f_{\ce{WS2}}$. Here, almost all excitons experience an intense electric field and consequently strongly couple to the GMR. In contrast, removing the monolayer from the hot-spot regions (anti-patterned case) causes the Rabi splitting to rapidly reduce with decreasing $f_{\ce{WS2}}$ (Fig.~\ref{fig:patterned}e). At $f_{\ce{WS2}} \approx 0.4$ the spectrum converges to a single peak close to the bare exciton energy $\hbar\omega_{X,\text{A}}$. This occurs because the remaining monolayer resides exclusively in weak-field regions, as captured in the spectrum at $f_{\ce{WS2}} = 0.24$ (black dashed line). In both structures, all features in the absorption spectrum disappear for $f_{\ce{WS2}} \to 0$, corresponding to the complete removal of the TMD. The spatial patterning of TMD monolayers on a PhC slab enables precise control over the relative contributions of WC and SC excitons to optical spectra. Additionally, this facilitates the design of ultra-compact open cavities where all excitons are strongly coupled, eliminating the background response of WC excitons. This mimics the behavior of closed cavities without the bulkiness of conventional DBR-based mirrors. Furthermore, this concept offers the capability to simultaneously engineer the Purcell effect and the strong coupling regime on a sub-$100$\,nm scale within a single photonic device. 

\section{Controlling exciton-light coupling with the photonic crystal geometry}

\begin{figure}[!ht]
    \centering
    \includegraphics[width=0.85\textwidth]{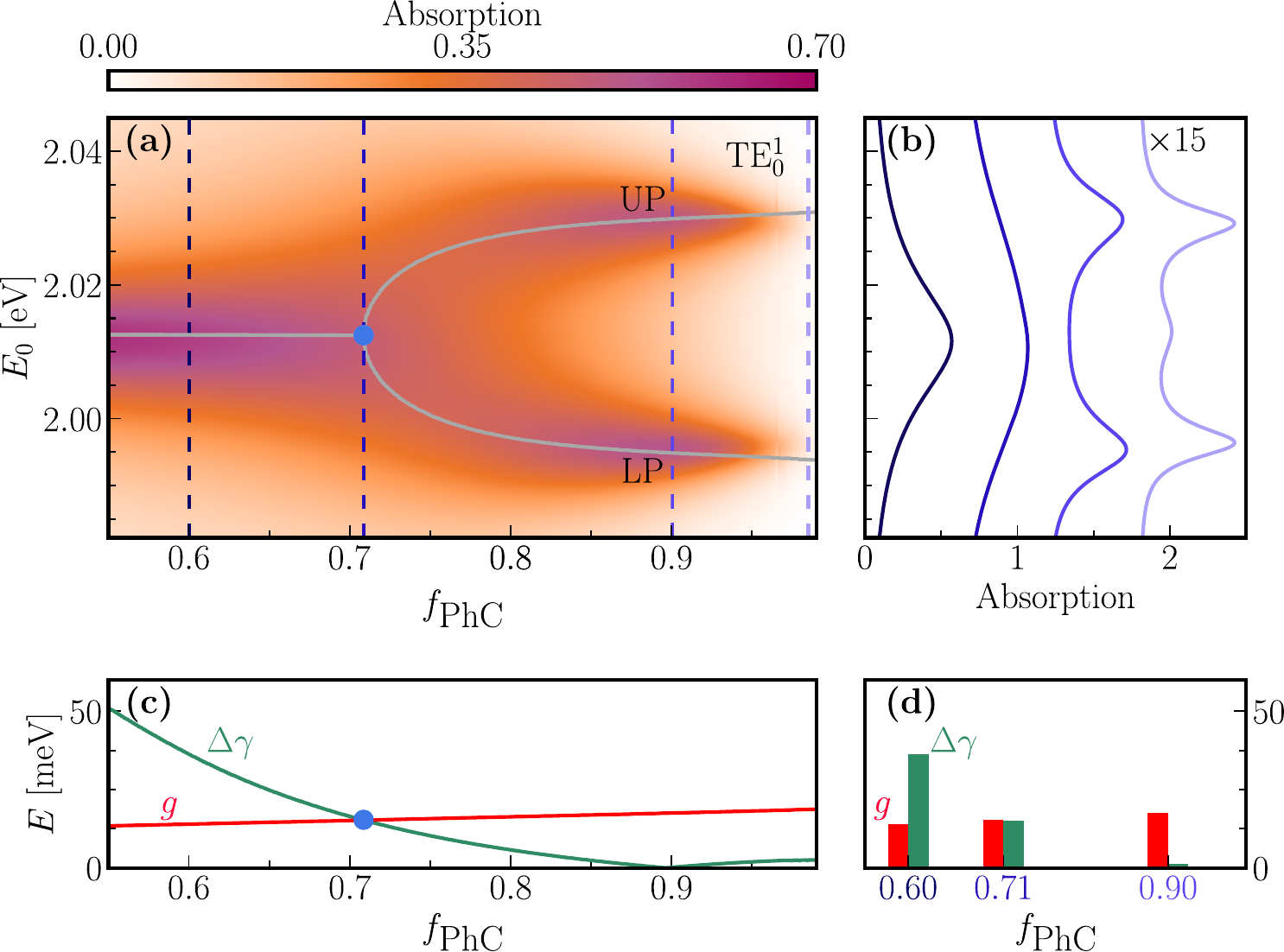}
    \caption{\textbf{(a)} Absorption spectra at normal incidence of the PhC/WS$_2$ structure versus $f_{\text{PhC}}$, and energy $E_0$. The PhC dimensions are scaled to maintain the resonance condition between the A-exciton and the TE$_0^1$ GMR. Solid gray lines indicate the polariton energies calculated via Eqs.~\ref{eq:g} and \ref{eq:polariton_energy}, with the exceptional point indicated by the blue dot. \textbf{(b)} Absorption spectra indicated by the vertical dashed lines in (a) using the same color code, offset by $0$, $0.6$, $1.2$ and $1.8$ respectively. \textbf{(c)} Comparison of the coupling strength $g$ (red line, Eq.~\ref{eq:g}) and the linewidth difference $\Delta \gamma = |\gamma_\text{GMR} - \gamma_X|/2$ (green line, extracted from RCWA spectra). \textbf{(d)} Values of $g$ and $\Delta \gamma$ for the three of the spectra shown in (b)}.
    \label{fig:ff_dependence}
\end{figure}

Finally, we investigate the strong coupling between the A-exciton resonance and the TE$_0^1$ GMR as a function of the scale-invariant filling factor, $f_{\text{PhC}}$. Since we want to explore the coupling over a range of large filling factors and use a single mode approximation, we switch from the TE$_0^2$ mode to the TE$_0^1$ mode. This ensures that the GMR is spectrally isolated from other modes for the PhC parameters explored in this section. Note that we scale the PhC geometry to maintain resonance between the TE$_0^1$ mode and the A exciton, details are provided in the SI. Varying the filling factor changes the linewidth and the in-plane electric field distribution of the GMR, thereby modifying the Rabi splitting, as shown in Fig.~\ref{fig:ff_dependence}a. For $f_{\text{PhC}}<0.71$ (exceptional point \cite{miri2019exceptional} indicated by the blue dot), the system lies in the weak coupling regime with a single peak in the absorption spectra. Above this point, this single peak splits indicating the strong coupling regime. The polariton energies and the exceptional point are accurately reproduced by the effective slab model (gray solid lines). 

Figure~\ref{fig:ff_dependence}b illustrates the transition between weak and strong coupling regimes through four representative spectra at $f_{\text{PhC}} = 0.6,0.71,0.9,0.985$. Below the exceptional point, a single peak is observed; above it, two distinct peaks emerge with a faint third peak appearing at high $f_{\text{PhC}}$. At the exceptional point itself ($f_{\text{PhC}} = 0.71$), the strong coupling condition is met, but the three peaks cannot be spectrally resolved, resulting in a single broadened peak in the absorption spectrum. Further discussion about the criteria and mathematical conditions used to determine the coupling regime can be found in the SI. The spectral position of the exceptional point is determined by the relation between the coupling strength $g$ and the difference in linewidths of the A exciton resonance and the GMR, $\Delta\gamma = |\gamma_\text{GMR} - \gamma_X|/2$, as shown in Fig.\ref{fig:ff_dependence}c. Consequently, the spectra in Fig.~\ref{fig:ff_dependence}b can be understood by comparing $g$ to $\Delta \gamma$ as shown in Fig.~\ref{fig:ff_dependence}d. For $f_{\text{PhC}} = 0.6$ the system is deep in the weak coupling regime, as $\Delta \gamma > g$, whereas for $f_{\text{PhC}} = 0.9$  the system is well inside the strong coupling regime, as $\Delta \gamma < g$. Crucially, since the coupling strength is approximately constant within the range shown here, the GMR linewidth is the dominant parameter driving the observed change in the Rabi splitting. 

\section{Conclusion}
In this work, we have identified the physical mechanisms governing strong light-matter coupling in systems composed of 1D PhC slabs and 2D semiconductors. Using a combination of rigorous coupled wave analysis, coupled-mode theory and an effective slab waveguide model, we clarify the origin of the three-peak structure in absorption spectra of strongly coupled open systems. Alongside the upper and lower polariton branches, which can also be observed in regular Fabry-P\'erot cavities, there is a third peak that can be attributed to weakly coupled excitons lying outside electromagnetic field hot spots in the PhC unit cell. The difference between these weakly coupled excitons and guided polaritons can be interpreted as a consequence of the varying in-plane electric field profiles and the strong localization within the unit cell. Alternatively, this distinction can be viewed as a consequence of zone-folding the slab waveguide polariton dispersions onto the bright exciton resonance due to the grating periodicity. Furthermore, by tailoring PhC and 2D semiconductor geometries, we modify the Rabi splitting and the relative contributions of strongly and weakly coupled excitons appearing in the optical spectra. Overall, our results can be of key importance for designing practical ultra-thin photonic devices that combine weak and strong coupling physics with 2D semiconductors on a sub-micron scale. 

\begin{acknowledgement}
The authors acknowledge funding from the Deutsche Forschungsgemeinschaft (DFG) via the regular project 524612380.
\end{acknowledgement}

\section{Disclosures}
The authors declare no conflicts of interest.

\section{Data Availablitiy}
Data underlying the results presented in this paper are not publicly available at this time but may be obtained from the authors upon reasonable request.

\begin{suppinfo}
The Supporting Information contains further details on the RCWA and CMT calculations, along with discussions of the waveguide polariton extended-zone scheme, effective medium theory for PhCs, strong coupling conditions, and the scaling of PhC parameters for the filling-factor studies.
\end{suppinfo}





\bibliography{bib}
\end{document}


\section{Simulation details}

All simulations shown in this work were performed using rigorous coupled wave analysis (RCWA), a semi-analytical method ideal for efficiently modeling layered structures with one- or two-dimensional in-plane periodicity \cite{whittaker1999scattering}. The method works by expanding the periodic permittivity and electromagnetic fields of each layer into a Fourier series. This approach transforms Maxwell's equations into a matrix eigenvalue problem for each layer, which is solved to find the characteristic modes. Finally, a scattering matrix formalism is used to enforce the boundary conditions of the tangential fields at each interface, leading to the reflection and transmission of the entire device \cite{rumpf2011improved}. The room temperature refractive index of \ce{WS2} was taken from \citet{li2014measurement}, where a multi-Lorentzian oscillator model was fitted to experimental data. Table 1 lists all RCWA simulation parameters for the presented calculations. Note that we used  $7\times1$ spatial harmonics and $2048 \times 1$ real space grid points to capture the photonic crystal (PhC) unit cell.
\begin{table}[ht!]
    \centering
    \begin{tabular}{|c|c|c|c|c|c|}\hline
        \centering
        Figure  & $E_0$ [eV] & $\beta$ [nm$^{-1}$] & $h$ [nm] & $\Lambda$ [nm] & $w$ [nm] \\
        \hline
        1b & $0.124 - 2.6$ & $0$ & $0-250$ & $521$ & $52.1$ \\
        1d & $0.124 - 2.6$ & $0-12.0$ & $75$ & $521$ & $52.1$ \\
        1e-g & $1.051,$ $1.788,$ $1.950$ & $0$ & $120$ & $521$ & $52.1$ \\
        \hline
        2e & $1.982 - 2.048$ & $0-6.0$ & $85$ & $521$ & $52.1$ \\
        \hline
        3a & $1.982 - 2.048$ & $0$ & $62-78$ & $521$ & $52.1$ \\
        \hline
        4d,e & $1.971-2.058$ & $0$ & $70.3$ & $521$ & $52.1$ \\
        \hline
        5a & $1.959-2.052$ & $0$ & $40.04-76.55$ & $296.29-566.49$ & $29.63-56.65$ \\
        \hline
    \end{tabular}
    \caption{Overview of RCWA simulation parameters for all plots shown in the main text.}
    \label{tab:RCWA_overview}
\end{table}  

\section{Waveguide polaritons in the extended-zone scheme}

\begin{figure}[b!]
    \centering
    \includegraphics[width=0.9\linewidth]{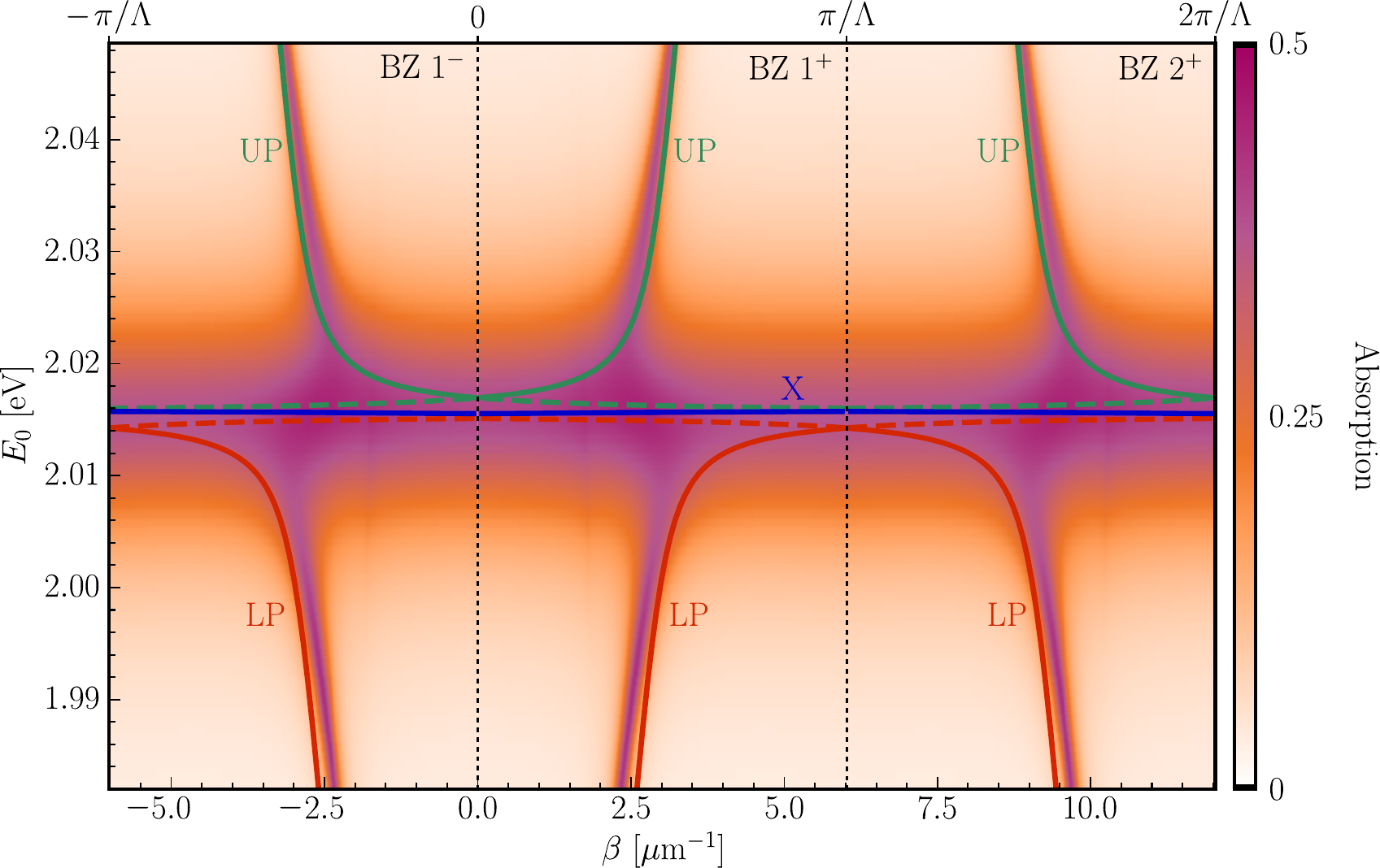}
    \caption{The extended Brillouin zone (BZ) scheme corresponding to Fig.~2e of the main text. The first BZ (BZ $1^\pm$: $\beta \in [- \pi/\Lambda,\pi/\Lambda]$) and the positive section of the second BZ (BZ $2^+$) are shown. Each part of the BZ features a set of upper (green lines) and lower (red lines) polariton branches as well as the flat weakly coupled exciton (blue line). Note that these dispersions extend outside the air lightcone.}
    \label{fig:extended_zone_folding}
\end{figure}

Figure~\ref{fig:extended_zone_folding} illustrates the extended-zone scheme, depicting how the upper (UP) and lower (LP) guided polariton dispersions of neighboring Brillouin zones (BZs) (BZ 1$^-$/BZ 2$^+$) extend into the positive half of the first BZ (BZ 1$^+$). These extensions of the LP/UP branches are shown as dashed green/red lines here and in Fig.~2e of the main text. An improved model that captures the photonic band gap would flatten these extended branches onto the weakly coupled exciton dispersion, while the UPs/LPs would merge at the BZ edges.

\section{Effective medium theory for photonic crystal slabs}

\begin{figure}[b!]
    \centering
    \includegraphics[width=0.6\linewidth]{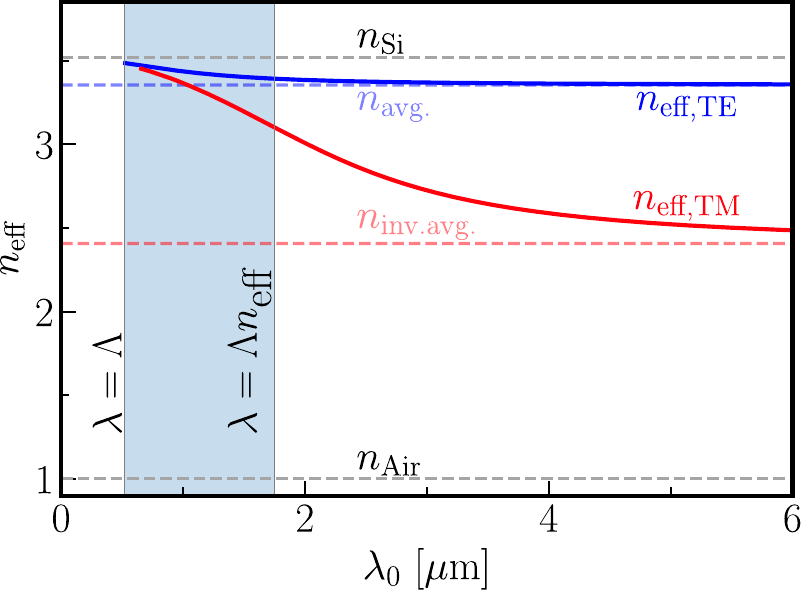}
    \caption{Effective refractive index for TE (blue solid line) and TM (red solid line) polarization as a function of the incident wavelength $\lambda_0$. The refractive indices of air and \ce{Si} are indicated by the grey dashed lines. The blue and red dashed lines indicate the averaged, $n_{\text{avg}}$, and inverse-averaged, $n_{\text{inv. avg}}$, effective refractive indices, respectively. The shaded area indicates the guided-mode regime as discussed for Fig.~1b in the main text.}
    \label{fig:fig_SIRytov}
\end{figure}

The effective slab theory presented in the main text relies on an accurate effective medium theory for the refractive index $n_{\text{eff}}$. Such theories are built on the assumption of a small thickness \cite{lalanne1997depth} or period \cite{hemmati2020applicability,rytov1955electromagnetic} with respect to the wavelength. The latter corresponds to Rytov's effective medium theory, which assumes an infinitely thick grating with a deep-subwavelength period. In general, separate effective refractive indices need to be introduced for TE and TM polarizations. In Fig.~\ref{fig:fig_SIRytov}, the effective refractive index from Rytov's theory (Eqs.~2 and 3 in Ref.~\citenum{hemmati2020applicability}) is shown for both polarizations as a function of the incident wavelength $\lambda_0$ for a fixed period of $521$$\,$nm. 
\\
Figure~\ref{fig:fig_SIRytov} reveals that $n_{\text{eff}}^{\text{TE}}$ varies between $n_{\text{Si}}$ at short wavelengths ($\lambda \approx \Lambda$) and the volume average $n_{\text{avg}} = \sqrt{f\varepsilon_{\text{Si}} + (1-f) \varepsilon_{\text{air}}}$ in the deep-subwavelength limit ($\lambda_0\gg\Lambda$). In contrast, $n_{\text{eff}}^{\text{TM}}$, while also limiting to the same short wavelength limit as $n_{\text{eff}}^{\text{TE}}$, approaches the volume average of the inverse permittivity at longer wavelengths $n_{\text{inv. avg}} = 1/\sqrt{f\varepsilon_{\text{Si}}^{-1} + (1-f) \varepsilon_{\text{Air}}^{-1}}$. In the guided-mode regime relevant for this work ($\Lambda n_{\text{eff}} > \lambda_0 > \Lambda$), we find that using the volume average for both polarizations is appropriate. This corresponds to the zeroth-order approximation used in the small-grating-depth expansion of Ref.~\citenum{lalanne1997depth}.

\section{Coupled mode theory of exciton–guided mode resonance interactions}

By transforming Eq.~4 of the main text into the frequency domain, we can extract the total scattering matrix of the device \cite{fan2003temporal,fan2002analysis}
\begin{equation}
    S = \begin{pmatrix} r && it \\ it && r\end{pmatrix} = \mathcal{C} + K \left(i \left( \omega I - \hbar\Omega \right) + \Gamma \right)^{-1} K^{T}. 
    \label{eq:s}
\end{equation}
For the three-resonator two-port system we are interested in, we obtain the resulting reflection/transmission coefficients
\begin{equation}
    \begin{split}
        S_{11} = r &= |r_d| + \frac{\gamma_\text{GMR} \eta_X \eta_{\tilde{X}} + \gamma_X \left(\eta_\text{GMR} \eta_{\tilde{X}} + g^2\right) - \gamma_\text{GMR} \gamma_X \eta_{\tilde{X}}}{\gamma_{\text{GMR}}\gamma_{X} \eta_{\tilde{X}} - \eta_\text{GMR} \eta_X \eta_{\tilde{X}} + \eta_{X}g^2}\\
        S_{12} = it &= i|t_d| \pm \frac{\gamma_\text{GMR} \eta_X \eta_{\tilde{X}} + \gamma_X \left(\eta_\text{GMR} \eta_{\tilde{X}} + g^2\right) - \gamma_\text{GMR} \gamma_X \eta_{\tilde{X}}}{\gamma_{\text{GMR}}\gamma_{X} \eta_{\tilde{X}} - \eta_\text{GMR} \eta_X \eta_{\tilde{X}} + \eta_{X}g^2}, \label{eq:s2}
    \end{split}
\end{equation}
where 
\begin{equation}
    \eta_j(\omega) = i\hbar(\omega_j - \omega) - \gamma_{j,\text{r}} - \gamma_{j,\text{nr}},
\end{equation}
with the mode energy, radiative and non-radiative decay rates, $\hbar \omega_j$, $\gamma_{j,\text{r}}$ and $\gamma_{j,\text{nr}}$, respectively. The scattering matrix of the background process has the form:
\begin{equation}
    \mathcal{C} = e^{i\phi}\begin{pmatrix} r_d & it_d \\ it_d & r_d\end{pmatrix},
\end{equation}
where the phase $\phi$ can be set to zero by appropriately defining the origin of the coordinate system. Linear optical spectra can then be calculated from Eqs.~\ref{eq:s2}. In particular, the absorption is obtained using energy conservation, $A = 1 - |r|^2 - |t|^2$. Additionally, the eigenvalues of the scattering matrix in Eq. \ref{eq:s} give the polariton energies, see Eq.~(7) of the main text.

\begin{figure}[t!]
    \centering
    \includegraphics[width=0.7\linewidth]{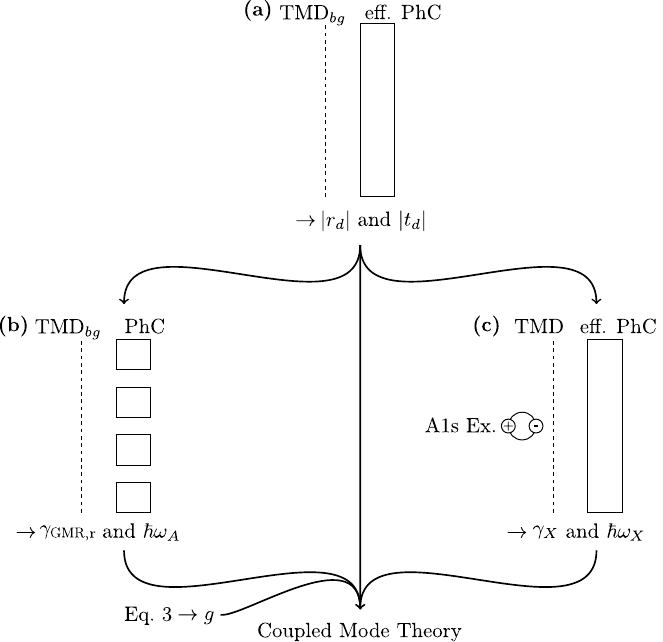}
    \caption{Schematic illustration of the different RCWA simulation configurations used to extract all the necessary parameters for the coupled mode theory. (a) indicates the calculation of the background scattering matrix by considering  a dielectric slab with the background permittivity of \ce{WS2} on top of a uniform dielectric slab with the effective refractive index $n_{eff}$. The latter captures the background slab Fabry-P\"erot modes of the grating. (b) The energy and radiative decay rate of the bare guided mode resonance can be extracted from RCWA simulations, where the impact of the TMD monolayer background is considered. (c) Likewise, the energy and total decay rate of the A exciton resonance can be extracted by simulating the \ce{WS2} monolayer on top of an effective dielectric slab with the refractive index $n_{\text{eff}}$.}
    \label{fig:CMT_Roadmap}
\end{figure}

To accurately model the linear optical response with the coupled mode theory, it is necessary to extract the relevant material-specific parameters: $\hbar\omega_X$, $\hbar\omega_\text{GMR}$, $\gamma_X$, $\gamma_\text{GMR}$, $r_d$, and $it_d$ from RCWA simulations. Rather than treating the coupling $g$ as a fitting parameter, we instead calculate it using Eq.~3 of the main text combined with electric field calculated using the effective slab waveguide model. The direct process $\mathcal{C}$ is determined analytically by removing the resonances of the structure leading to a stack of two dielectric slabs, i.e., a dielectric slab with the background permittivity of \ce{WS2} on top of a uniform dielectric slab with the effective refractive index $n_{\text{eff}}$ (Fig.~\ref{fig:CMT_Roadmap}a). This yields $r_d(\omega)$ and $it_d(\omega)$. The spectral position and linewidth of the bare leaky guided mode resonance can be extracted from RCWA simulations, where the exciton resonances of the TMD monolayer are removed, i.e., a dielectric slab with the background permittivity of \ce{WS2} on top of the 1D PhC with the refractive index $n_{\text{Si}}$ (Fig.~\ref{fig:CMT_Roadmap}b). Analogously, the spectral position and linewidth of the A exciton can be extracted by simulating the \ce{WS2} monolayer on top of an effective dielectric slab with the refractive index $n_{\text{eff}}$ (Fig.~\ref{fig:CMT_Roadmap}c). Consequently, the Purcell effect and Lamb shift due to the change in dielectric environment of the TMD monolayer (compared to suspension in air) are automatically included in $\gamma_{X,\text{r}}$ and $\hbar \omega_X$. \\
We assume the oscillator strength in Eq.~3 to be an intrinsic value, i.e., independent of the dielectric environment of the TMD monolayer. Therefore, we can calculate it from the radiative decay rate extracted from RCWA simulations of the TMD monolayer in vacuum with the following equation:
\begin{equation}
    f = \frac{4 \varepsilon_0 n_\text{bg} m_\text{e} c_0}{\hbar e^2}\gamma_{X,\text{r}},
\end{equation}
where we can set $n_\text{bg}=1$. The radiative and non-radiative decay rates of the exciton can be extracted by the following two equations \cite{kira2006many}:
\begin{equation}
    \begin{split}
        \text{FWHM} &= 2 \left(\gamma_{X,\text{r}}+ \gamma_{X,\text{nr}}\right), \\
        A_\text{max} &= \frac{2 \gamma_{X,\text{r}} \gamma_{X,\text{nr}}}{\left(\gamma_{X,\text{r}} + \gamma_{X,\text{nr}}\right)^2},
    \end{split}
\end{equation}
provided the exciton resonance is well separated from neighboring resonances. Here FWHM is the full-width at half-maximum of the peak in absorption while $A_\text{max}$ is the maximum absorption. \\
Lastly the mode volume $L_\text{mode}$ is defined by \cite{ciers2017propagating}
\begin{equation}
    L_\text{mode} = 2 \frac{\int \varepsilon(z) |E(z)|^2 \mathcal{d} z}{\max{\varepsilon(z) |E(z)|^2}},
\end{equation}
where $\varepsilon(z)$ is the spatially dependent relative permittivity and $E(z)$ is the electric field.

\section{Conditions for the strong coupling regime}
From Eq.~7 of the main text, we calculate the Rabi-splitting $\hbar\Omega_R$ from the extracted linewidths $\gamma_A$ and $\gamma_X$, energies $\hbar \omega_A$ and $\hbar \omega_X$, and the calculated coupling strength $g$ (Eq.~3 of the main text). We numerically demonstrate that the strong coupling condition (at resonance $E_X = E_A$) is met:\cite{schneider2018two}
\begin{equation}
    \begin{split}
        1 &> \frac{|\gamma_X - \gamma_A|}{2g} \approx 0.05,\\
        1 &> \frac{\gamma_X + \gamma_A}{2\hbar\Omega_R} \approx 0.15.
    \end{split}
\end{equation}
The first expression corresponds to the theoretical condition for strong coupling, for which the expression under the square root in Eq.~7 of the main text is positive at zero detuning. However, to resolve the polariton peaks in absorption spectra the second condition must be fulfilled, i.e., the Rabi splitting has to be larger than the combined linewidths of the exciton and photon mode.

\newpage
\section{Scaling photonic crystals}
\begin{figure}[ht!]
    \centering
    \includegraphics[width=0.9\textwidth]{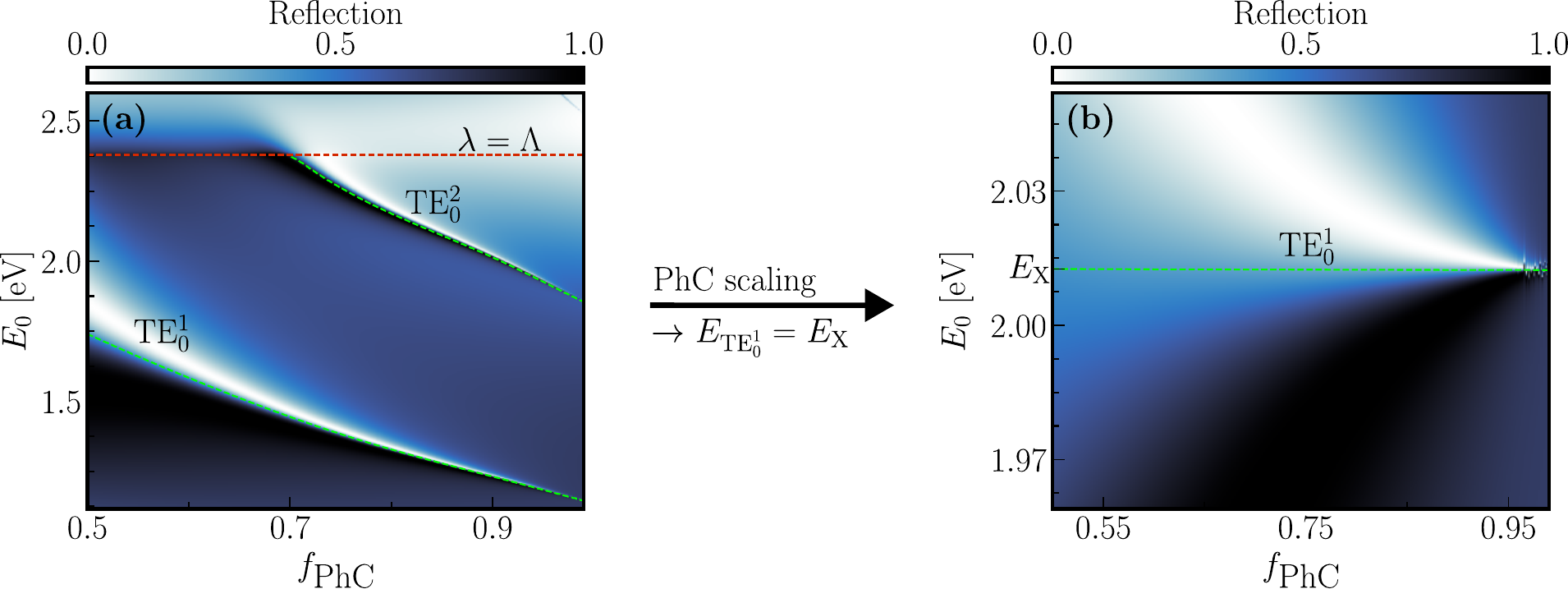}
    \caption{\textbf{(a)} Normal incidence reflection spectra (TE) of Fig.\,5 in the main text without the TMD and before scaling of the PhC. Both the spectral position and linewidth of the TE$_0^1$/TE$_0^2$ modes of the PhC (green dashed lines) decrease with the increasing filling factor. The dashed red line indicates the transition between the diffractive and guided mode regimes ($\lambda = \Lambda$). The TE$_0^2$ mode crosses into the diffractive regime for $f \approx0.7$. \textbf{(b)} The same reflection spectra after scaling of the PhC, such that the resonance condition between the A exciton and the TE$_0^1$ mode is satisfied. The spectral position of the PhC mode is now constant, while the linewidth still increases for decreasing filling factors.} 
    \label{fig:SIPhotonicMode}
\end{figure}

Figure~\ref{fig:SIPhotonicMode} shows the TE$_0^1$ and TE$_0^2$ mode before and after the scaling of the PhC over the filling factor $f_{\text{PhC}}$. In Fig.~\ref{fig:SIPhotonicMode}a, the spectral position of the leaky guided modes redshifts with increasing filling factor. Likewise, the linewidth narrows for larger filling factors, and completely vanishes as $f_{\text{PhC}} \rightarrow 1$. The transition between the guided-mode and refractive regime is visible in the reflection spectra (Fig.~\ref{fig:SIPhotonicMode}a) at $E_0 \approx 2.4\,$eV (dashed red line). This transition only depends on the period, $\Lambda$, as discussed in Fig.~1b of the main text. As the TE$_0^2$ mode crosses into the diffractive regime for $f \approx 0.7$ it is not suitable for the single-optical-mode approximation employed in Fig.~5. Therefore, we chose the TE$_0^1$ mode, which lies in the guided-mode regime for all $f_{\text{PhC}}$ shown here. 
\newpage

\bibliography{bib}